\documentclass[12pt]{article}

\usepackage{psfig}

\newlength{\figwidth}
\newcommand{\eabe} {\begin{eqnarray}}
\newcommand{\eaen} {\end{eqnarray}}
\newcommand{\eqbe} {\begin{equation}}
\newcommand{\eqen} {\end{equation}}

\newcommand{\mrm} {\mathrm}
\newcommand{\srm}[1] {_{\mathrm{#1}}}
\newcommand{\ol} {\overline}
\renewcommand{\ln} {\mrm {ln}}
\newcommand{\Nc} {{N\srm c}}
\newcommand{\ra} {\rightarrow}

\newcommand{\bibl}[5]
	{#1, {\it #2} {\bf #3} (#4) #5}
\newcommand{\ibid}[3]
	{{\it ibid.} {\bf #1} (#2) #3}
\newcommand{\anti}[1] {${ \ol \mrm {#1} }$}
\newcommand{\pair}[1] {${\mrm {#1 \ol #1} }$}

\begin{document}

\begin{titlepage}
\begin{flushright}
 LU TP 98-25\\
 November 1998
\end{flushright}
\vspace{25mm}
\begin{center}
  \Large
  {\bf Using Two-jet Events to Understand Hadronization} \\
  \vspace{12mm}
  \normalsize
  Patrik Ed\'en\footnote{e-mail patrik@thep.lu.se}, G\"osta Gustafson\footnote{e-mail gosta@thep.lu.se}\\
  Department of Theoretical Physics\\
  Lund University\\
\end{center}
\vspace{5cm} 
{\bf Abstract:} \\ 
While the hard phase of the strong interaction is well described by
perturbative QCD, the soft hadronization phase is less understood.
Benefiting from the high statistics from $e^+e^-$ experiments at the
${\mrm Z}^0$ resonance, it is possible to impose strong two-jet cuts
on the data without loosing the statistical significance. In these
events perturbative activity is suppressed and hadronization effects
can be more prominent. We show that after proper event cuts a set of
observables are sensitive to differences in the hadronization
models. These observables can thus be important tools for a more
detailed study of the hadronization mechanism.

\end{titlepage}

\section{Introduction}\label{sec:intro}
High energy reactions like $e^+e^- \rightarrow$ hadrons are usually
described in terms of two phases, an initial hard perturbative phase
formulated as a parton cascade followed by a soft non-perturbative
hadronization phase, for which phenomenological models are needed. The
first phase can be calculated from perturbative QCD, but as exact
perturbative results in most cases are available only to second order
approximate schemes have to be used. In the large $\Nc$ limit planar
diagrams dominate~\cite{planar}, and for  $e^+e^-$-annihilation, descriptions in terms
of parton cascades have been quite successful. 

Global features like shape variables
(thrust, sphericity, oblateness, major, minor etc.) and inclusive
particle spectra in e.g.\ $x_F$ and $p_\perp$, reflect mainly the parton
distribution in the initial perturbative phase. These observables can be
well described assuming that the hadron distribution closely follows the
distribution of partons (local parton hadron duality, LPHD~\cite{lphd}), provided a locally invariant cut-off is applied to the cascade. One example is the observed peak position of the distribution in the variable $\xi = \ln (1/x_F)$, which  for most hadrons agrees with parton level results, using a virtuality cut-off $Q_0$ which is larger for heavier hadrons. Also in some more phenomenological models, e.g.\ in the cluster fragmentation model~\cite{cluster} and the UCLA model~\cite{ucla} the hadronization properties are to a large extent determined by the hadron masses.

A more detailed description of the final hadronic state, including e.g.\
flavour, the baryon to meson ratio, spin and polarization as well as
correlations (including Bose-Einstein correlations), depends also on the
non-perturbative properties of QCD, including the structure of the vacuum
condensate. These features cannot be obtained from perturbative
calculations. Although some of them may in the future be accessible e.g.\
from lattice calculations, at present it is unavoidable that a model
attempting to give a detailed description of the hadronization process
must contain a significant number of phenomenological parameters.

To gain insight in the properties of QCD in the soft region and the
vacuum structure it is essential to isolate the hadronization process
from the perturbative cascade. This is however not an easy problem. At
the end of the cascade the running $\alpha_s$ becomes relatively large.
The result depends on a necessary infrared cut-off and nonplanar diagrams
and interference effects may be important. This implies that this phase
is more uncertain and therefore difficult to isolate. It is also possible
that the separation between the perturbative and non-perturbative regions
cannot be well defined. As discussed in ref~\cite{conny} a change in the
perturbative cut-off can to a large extent be compensated by a
modification of the hadronization parameters. The transition region
between the two domains may also show interesting coherence effects~\cite{screw}.

In this paper we will study ways to isolate high energy systems with little
perturbative activity, in which the non-perturbative features ought to be
more prominent. With the very high statistics available from the LEP1 experiments, it is possible to impose event cuts which exclude a significant amount of perturbative activity, and still have remaining events numerous enough for detailed studies of hadronization.
It is possible to calculate purity and efficiency
measures for different event shape cuts. Our result is however that a simultaneous purity and efficiency close to 1 cannot be obtained; there is no cut that can be called optimal. Instead different cuts are suitable for a study of different features of the hadronization process.

The outline of this paper is as follows: In section~\ref{sec:cuts} we study a set of different event cuts. In section~\ref{sec:obs} we
study some observables which have been proposed for a separation of
different hadronization mechanisms. In section~\ref{sec:models} we discuss a few models with different properties and in section~\ref{sec:results} their predictions for the considered observables. Conclusions are given in section~\ref{sec:concl}.

Much of the discussion in this paper will concern different aspects of transverse momenta. These will be denoted $k_\perp$ for partons and $p_\perp$ for hadrons. We will also discuss a sum of hadron $p_\perp$, which we will call $Q_\perp$. 

\section{Cuts}\label{sec:cuts}

We will here discuss a set of event shape cuts suitable to extract events with little perturbative activity. The performance of the cuts are investigated using Monte Carlo simulations, where the underlying parton state is known.

For simulated events we will here with ``little perturbative activity'' mean events where all gluon emissions occur at a $k_\perp$ below some $k_{\perp0}$. (In the analysis presented below we will take $k_{\perp0}=2$ GeV.) The reason for choosing this definition, rather than e.g.\ a cut on the sum of $E_\perp$ for all gluons, is that one hard gluon gives rise to a distinct three jet structure, which introduces long range correlations in azimuthal angle. Such a gluon also increases the phase space for softer ones, which enhances its effects on the final state.

Through further emissions, the energy of one gluon is distributed among its emission products. As this rather enhances than reduces the importance of the perturbative phase, we require $k_\perp$ to be smaller than $k_{\perp0}$ for each emission, and not for the final partons at the end of the cascade.

We examine the performance of the cuts in terms of purity and efficiency. Purity is defined as the rate of events where the highest $k_\perp$ for a gluon emission, $k_{\perp\mrm{max}}$, is below $k_{\perp0}$. If the purity is low, perturbative gluons may shadow many hadronization effects. Efficiency is defined as the acceptance rate among the desired events with $k_{\perp\mrm{max}}$$<$$k_{\perp0}$. If the efficiency is low, the sample might be biased, and the interpretation of the results is more difficult. Low efficiency also implies low statistics. 

The cuts are applied to hadronic final states from ${\mrm Z}^0$ decays, in events generated by the \textsc{Ariadne} and \textsc{Jetset} MC~\cite{ariadne,jetset}. \textsc{Ariadne} is an implementation of the colour dipole formalism for a QCD cascade~\cite{cdm}. This cascade is convenient to use in the present analysis, as it is ordered in  $k_\perp$, which implies that $k_{\perp\mrm{max}}$ of the cascade is easily extracted as the $k_\perp$ of the first gluon emission. \textsc{Jetset} is a MC for the Lund string fragmentation model~\cite{lundstring}, which in general gives a good description of $e^+e^-$ data.

The considered two-jet measures are 
\begin{itemize}
\item $1-T$ ($T$=Thrust=$\max_{\ol n_T}\frac{ \sum|\ol n_T \cdot \ol p|}{ \sum|p|}$).
\item $M$ ($M$=Major=$\max_{\ol n_M \perp  {\ol n_T}}\frac{ \sum|\ol n_M \cdot \ol p|}{ \sum|p|}$). 
\item $k_{\perp \mrm{clus}}$= transverse momentum of three-jet configuration found using $k_\perp$-based cluster method (we have chosen the \textsc{Diclus} algorithm~\cite{diclus}).
\item Sphericity $S$, which measures the sum of $p_\perp^2$ w.r.t.\ the axis minimizing this sum.
\item ``Linearized sphericity'' $L$, which is similar to sphericity, but linear in particle momenta.
\item Multiplicity.
\end{itemize}
Sphericity $S$ is determined by the largest eigenvalue $\lambda\srm{max}$ to the sphericity tensor 
\eqbe S^{\alpha\beta}\equiv \frac{\sum_ip_i^\alpha p_i^\beta}{\sum_i|{\mathbf p}_i|^2},\eqen
through the definition $S=\frac3 2(1-\lambda\srm{max})$. Thus $S\approx 0$ for a two-jet event and $S\approx 1$ for an isotropic event.
Sphericity is quadratic in particle momenta, which implies that the value is changed if a particle is split in two collinear ones. A linear correspondence to sphericity, suitable for perturbative calculations~\cite{linsph}, is obtained by a modified tensor 
\eqbe L^{\alpha\beta}\equiv  \frac{\sum_ip_i^\alpha p_i^\beta/|{\mathbf p}_i|}{\sum_i|{\mathbf p}_i|}.
\eqen
The quantity $L=\frac3 2(1-\lambda^{(L)}\srm{max})$ thus measures the summed $p_\perp^2/|{\mathbf p}|$ w.r.t.\ the event axis. In this paper we will use the terminology suggested by Sj\"ostrand, and call this measure linearized sphericity.

The  event shape cuts in thrust, sphericity and jet-$k_\perp$ are widely used to obtain two-jet events. Apart from these quantities, we have chosen to investigate two measures which can be expected to be closely related to the maximal $k_\perp$ of the parton cascade. One is major, which is given by the maximal summed $|{\mathbf p}|$ component orthogonal to the thrust axis, and the other is linearized sphericity discussed above. The cut in multiplicity is used partly to ensure that the more carefully designed cut indeed perform better. Nevertheless it enhances the rate of events with little perturbative activity, as the emitted gluons in general give rise to high multiplicities. One advantage of this cut, which makes it worth consideration, is that $p_\perp$ observables are left relatively unbiased.

\begin{figure}[tb]
  \begin{center}  \hbox{ \vbox{
	\mbox{\psfig{figure=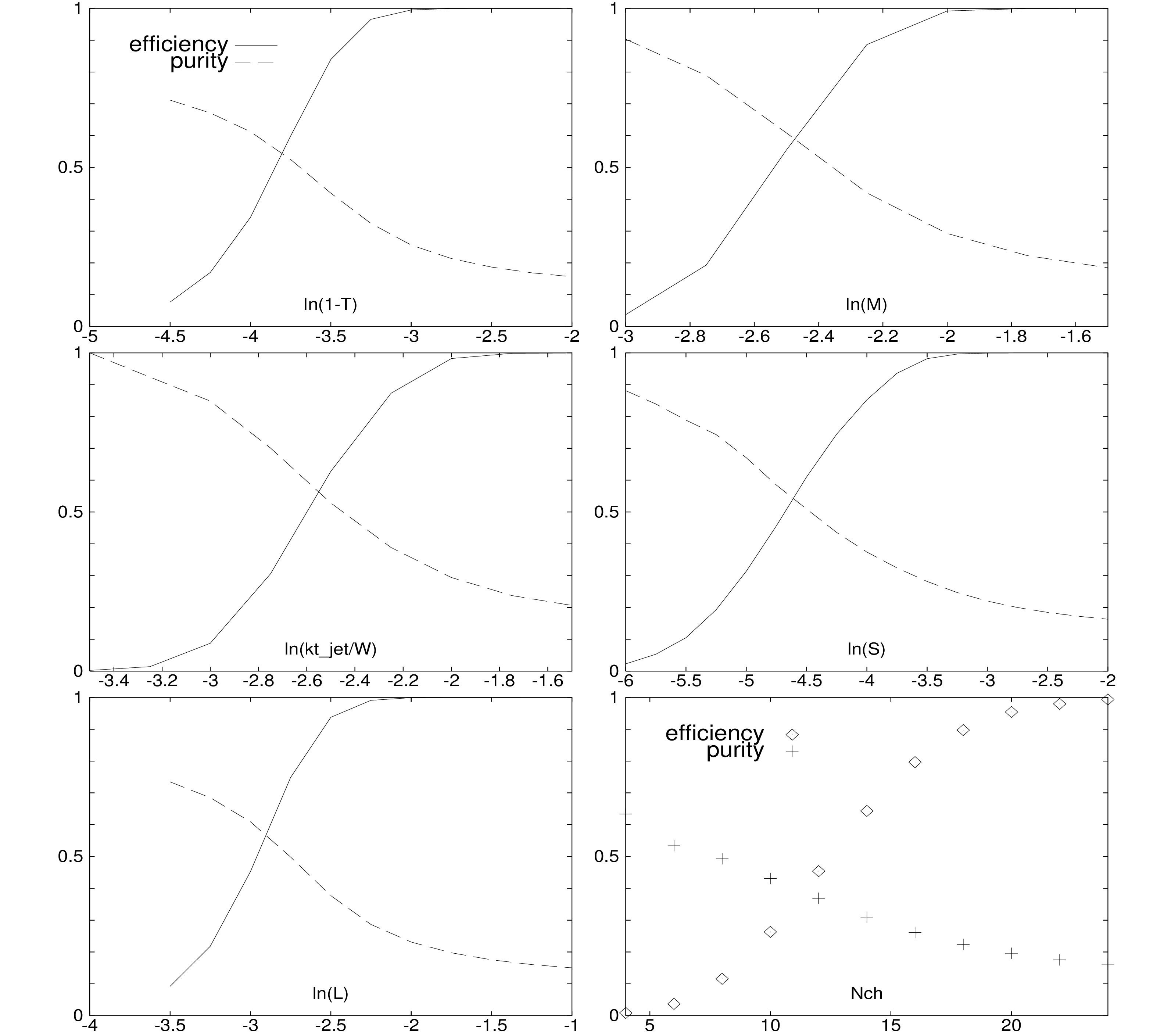,width=\textwidth}}
   }  }
  \end{center}  \caption{\em Purity and efficiency for different two-jet cuts.}
  \label{f:cuts}
\end{figure}
Fig~\ref{f:cuts} shows efficiency and purity results for a $k_{\perp0}$$=2$ GeV for the simulated gluon emissions. Most two-jet cuts perform similarly, but it could be noted that the commonly used cut in $1-T$ not obviously is the most successful. For large rapidities of the emitted gluon jet, $1-T$ is approximately $4k_\perp^2/s$ and thus not very restrictive in $k_\perp$. In the central rapidity region, however, $1-T$ behaves as $k_\perp/\sqrt s$. For a study in this region, a cut in $1-T$ therefore performs very similarly to the other ones considered in Fig~\ref{f:cuts}.
The multiplicity cut performs noticeably worse than the others. This is expected, since it is not an cut explicitly in a $p_\perp$ variable.

We end this section by noting that the purity and efficiency measures are not very sensitive to the choice of cut. There is thus no optimal cut in all respects, and as we will see, the preferred cut instead depends on the observable under investigation.

\section{Observables}\label{sec:obs}
In this section we want to study a set of observables which depend on the
properties of the hadronization mechanism, but where this dependence is
masked by perturbative gluon emission. Thus we will not discuss here the
flavour composition of the hadrons, as this appears to be less affected
by gluon radiation. (We note, however, that this may not be completely
true, as there are experimental indications for a larger procution rate of $\Lambda$-particles in $\Upsilon$ decays~\cite{upsilon}, and also a larger rate of high-energy $\eta$-particles in gluon jets~\cite{etagluonjet}, than expected from available fragmentation models.) 

The observables discussed here are all related to transverse momentum. 
In many models $p_\perp$ of neighbouring hadrons are anti-correlated to a larger or lesser degree. This anti-correlation is counteracted by jet (or minijet) emission, which gives a positive correlation between hadrons close in rapidity, making the analysis of the soft $p_\perp$ generation very nontrivial.

\subsection{Inclusive $p_\perp$-distribution}\label{sec:obspt}

The models discussed in the following sections all have tunable
parameters for the soft $p_\perp$ generation. The values of these
parameters can be adjusted to reproduce the inclusive distribution of
the hadronic $p_\perp$. Thus this distribution acts as a constraint and
cannot be used to separate the different models. For this we have to use
more complicated observables, e.g.\ collective variables or
correlations.

\subsection{${\mathbf p}_\perp$ transfer}~\label{sec:obsQt}
One collective variable which can distinguish between different models
is the vector sum ${\mathbf Q}_\perp(y)$ of all hadrons with rapidity less
than a given rapidity $y$
\eqbe{\mathbf Q}_\perp(y)\equiv \sum_{y_i<y} {\mathbf p}_{\perp i}, \label{e:Qtdef} \eqen
which measures the $p_\perp$ transfer over the rapidity $y$.
(This variable has frequently been advocated by E. de Wolf~\cite{WolfQt}.)
For a
longitudinal phase space model constrained by overall momentum
conservation, $Q_\perp$ grows with the number of particles as a
random walk. Thus in this model $Q_\perp$ is larger in the central region
and larger in events with high multiplicity. In many fragmentation models, however, the ${\mathbf p}_\perp$ of a hadron is compensated locally in phase space. In such a model, $Q_\perp$ is limited and essentially independent of rapidity and multiplicity.

An essential point when using the $Q_\perp$ variable is the choice of axis. In an
experiment the direction of the initial \pair q pair is not known.
If the thrust axis is used to define the transverse directions,
$Q_\perp(y)$ is kinematically constrained to be equal to zero for $y=0$.
This definition is therefore very unsuitable, if one wants to use the
value of $Q_\perp$ in the central region to distinguish between different
models.

\begin{figure}[tb]
  \begin{center}  \hbox{ \vbox{
	\mbox{\psfig{figure=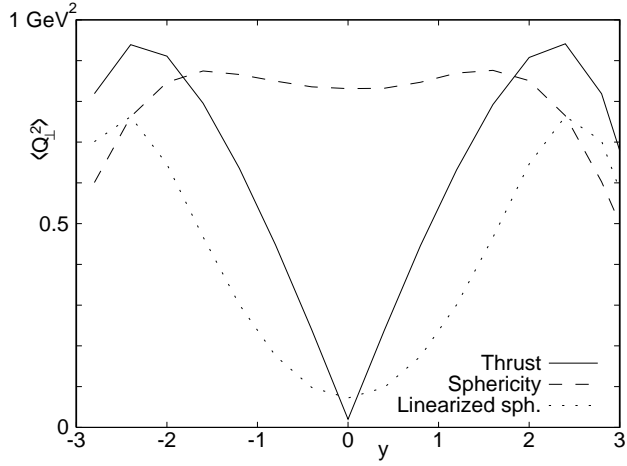,width=\figwidth}}
   }  }
  \end{center}  \caption{\em The properties of $\left<Q_\perp^2\right>$ in MC simulated events depend on the choice of axis. The thrust axis, which maximizes longitudinal momentum, by construction gives $Q_\perp=0$ when $y=0$. A similar behaviour is seen for Linearized sphericity. The conventional Sphericity axis, minimizing squared transverse momentum, gives the central rapidity plateau expected from the chosen string fragmentation model.}
  \label{f:qtaxes}
\end{figure} 
We have measured $Q_\perp$ w.r.t.\ different axes in MC generated events,
 using the \textsc{Jetset} default fragmentation scheme, where $p_\perp$ is compensated essentially by neighbouring particles in rapidity. This implies that we expect a rapidity plateau in $Q_\perp$, though decays of unstable hadrons and the perturbative activity present in the analysed two-jet events introduce corrections to this result.
As seen in Fig~\ref{f:qtaxes}, both the thrust axis and the linearized sphericity axis by construction create a significant dip  in $Q_\perp$ when $y=0$. The sphericity axis, on the other hand, reproduces the plateau extremely well.
 This should not be over-interpreted, for the reasons mentioned above, but it is clear that the sphericity axis is to prefer in a $Q_\perp$ analysis, and it will be used in the results presented below.

The collective quantity $Q_\perp$ is sensitive to correlations between neighbouring hadrons. It is also
possible to study directly the correlation in $p_\perp$ between two
hadrons separated by a rapidity interval $\delta y$
\eqbe \left<{\mathbf p}_\perp(y_1) \cdot {\mathbf p}_\perp(y_2=y_1 + \delta y)\right>,
\eqen
or the correlation between ${\mathbf Q}_\perp$ for two separated y-values
\eqbe \left<{\mathbf Q}_\perp(y_1) \cdot {\mathbf Q}_\perp(y_2=y_1 + \delta y)\right>.
\eqen
\begin{figure}[tb]
  \begin{center}  \hbox{ \vbox{
	\mbox{\psfig{figure=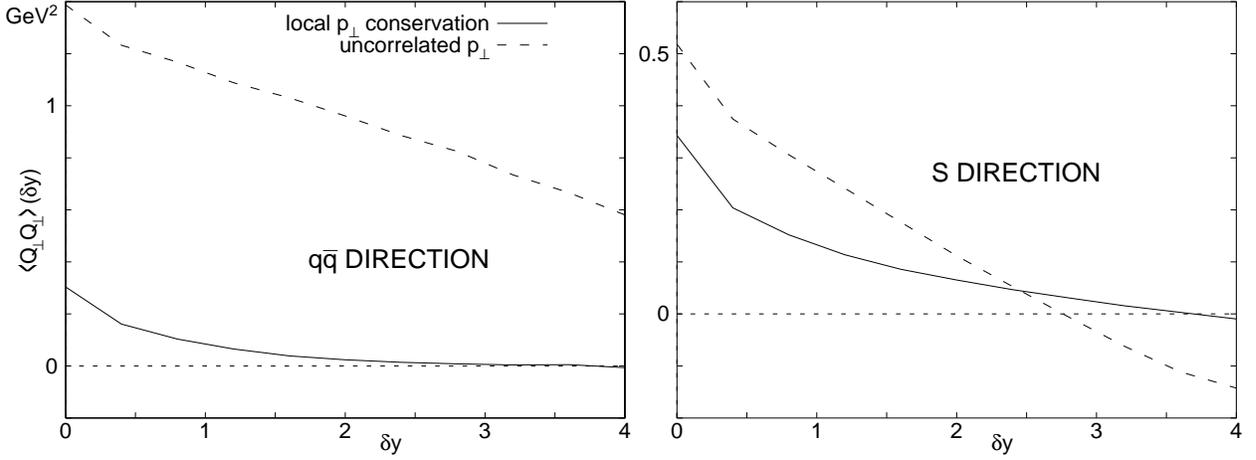,width=\textwidth}}
   }  }
  \end{center}  \caption {\em An explicit measurement of the correlation length $\delta y$ is distorted by the choice of axis. Two very different $p_\perp$ compensation models are compared when measured w.r.t.\ the original \pair q axis (left) and the sphericity axis (right). To emphasize the effects of axis choice, both the parton cascade and hadron decays have been turned off in the simulations.}
  \label{f:axisdelta}
\end{figure} 
The interpretation of these observables will however be hampered by the systematic effects related to the reconstruction of the event axis. In Fig~\ref{f:axisdelta} we compare $\left<{\mathbf Q}_\perp\cdot {\mathbf Q}_\perp\right>(\delta y)$ for two models, one with local $p_\perp$ conservation and one with essentially uncorrelated $p_\perp$. In the MC simulations, we have taken the liberty to turn off both the parton cascade and the hadron decays, in order to more clearly illustrate the effects of axis choice. The plots show the results obtained w.r.t.\ the original \pair q axis and the sphericity axis, respectively, and it is clear that the behavior of $\left<{\mathbf Q}_\perp\cdot {\mathbf Q}_\perp\right>(\delta y)$ is drastically changed. (Equally drastic changes occur if using the thrust axis or linearized sphericity axis.) Due to these effects of axis reconstruction in $e^+e^-$, a variation of $\delta y$ does not seem to provide much more information than is available from the observable $\left<Q_\perp^2\right>$. This orgument does {\em not} imply that the $\delta y$ correlation cannot be an interesting observable in deep inelastic scattering and hadron-hadron collisions, where the event axis is better known.

\subsection{Screwiness}\label{sec:obsscrew}
A particular correlation between rapidity and azimuthal angle
is determined by a variable called screwiness and introduced in ref~\cite{screw}.
Screwiness is defined by
\eqbe S(\omega) \equiv \frac1 {\left<N(y\srm{cut})\right>} \left<\left|\sum_{|y_i|<y\srm{cut}}\exp(i(\omega y_i - \phi_i))\right|^2\right>, \eqen
where $y_i$ and $\phi_i$ is the rapidity and azimuthal angle for particle $i$, and the sum runs over all central rapidity particles, specified by some $y\srm{cut}$. For large $\omega$ the sum describes a random walk along unit vectors in the complex plane, and the absolute square of the sum becomes the average multiplicity in the considered interval, $\left<N(y\srm{cut})\right>$. We have normalized $S(\omega)$ by this quantity, which implies that it is 1 for large $\omega$.

The screwiness measure is constructed to give a signal for a helix-like correlation between
$y$ and $\phi$. A motivation why such a correlation might be expected
is presented in ref~\cite{screw} and shortly described in the following section.

\section{Models}\label{sec:models}

The most studied hadronization models are based on string dynamics or
cluster fragmentation. In the cluster model implemented in the \textsc{Herwig} MC~\cite{herwig} the gluons split into \pair q pairs at the end of the cascade.
A quark is then combined with an adjacent antiquark to form a colourless
cluster, which normally decays to two (possibly unstable) hadrons. This
approach becomes less realistic and must be modified in events with
unusually low gluon radiation. In such events some clusters can get very
large masses, and in the \textsc{Herwig} MC these clusters decay in a stringlike
fashion. Consequently, if we select events with low gluon activity we
cut away those events for which \textsc{Herwig} is meant to work best. For this
reason we feel that it would not be meaningful to compare such an event
sample with \textsc{Herwig} simulations, and therefore we will instead in the
following study some different versions of string hadronization, and for
comparison also a model with a $p_\perp$ behaviour similar to a longitudinal phase space.

\subsection{String fragmentation}\label{sec:stringfrag}
In the Lund string model the confining colour field is assumed to behave
as a massless relativistic string starting at a quark, passing the
gluons in accordance with their order in colour, and ending at an
antiquark. Thus the gluons act as transverse excitations or kinks on the
string. The string breaks by the production of \pair q pairs (or
occasionally (qq)(\anti{qq}) systems) which combine to hadrons.

For the longitudinal momentum distribution it is assumed that any
breakup divides the string in two causally disconnected pieces, which
decay further independently of each other. Together with the assumption
of a central rapidity plateau this implies a unique result, the
symmetric Lund fragmentation model~\cite{symmetriclund,lundstring}. The result can be generated in an
iterative way with a splitting function of the form
\eqbe
f(z) \propto \frac{(1-z)^a}{z}\exp(-b M_\perp^2/z), \label{e:fz}
\eqen
where $M_\perp$ is the transverse mass of the generated hadron and $z$ is its fraction of jet energy plus longitudinal momentum.

The iterative procedure specifies an ordering of hadrons called rank. The original quark $\mrm{q}_0$ combines with an antiquark \anti{q}$_1$ from the \pair{q_1} breakup into the rank one meson. $\mrm{q}_1$ then forms the rank two meson together with the antiquark \anti{q}$_2$ from the next breakup, etc. Though the order in rank is not an observable, it is strongly correlated to the order in rapidity.

The string breakup by \pair q creation has been treated as a
tunneling process~\cite{tunnel, tunnelhyperbola}, in which the transverse momenta should be described
by a Gaussian with a width determined by the string tension $\kappa$:
\eqbe
\frac{\mrm dP}{\mrm dk_\perp^2} \propto \exp(-m_\perp^2/\sigma^2). \label{e:tunnel}
\eqen
Here  $m_\perp^2 = m^2 + k_\perp^2$ is the squared transverse mass of the quark or the antiquark, and from the tunneling mechanism the width should be given by $\sigma^2 = \kappa/\pi$. 

An essential property of the string fragmentation is infrared stability, meaning that a soft or
collinear gluon has only little influence on the string motion and
therefore also on the final hadronic state\
\footnote{To treat situations where a hadron obtains energy and momentum from several small string pieces is however not uniquely specified from the string dynamics. In the models studied below we will use the method developed by Sj\"ostrand~\protect\cite{gluonkinks} and implemented in the \textsc{Jetset} MC.}.
This feature makes the
result less sensitive to the cutoff for the partonic cascade, and a
change in the cutoff can be well compensated by a change in the hadronization parameters when looking at event shapes and inclusive observables for the total event sample~\cite{conny}.
Thus a larger cutoff is compensated by
a larger width $\sigma$ for the hadronization $k_\perp$. Some differences
may remain, however, as a gluon tends to give extra $k_\perp$ in the same
direction to neighbouring hadrons, while hadronization-$p_\perp$ tends to be in opposite directions, depending on the details in the fragmentation model.

In section~\ref{sec:obsQt} we discussed an observable called ${\mathbf Q}_\perp$, which is sensitive to the properties of $p_\perp$ compensation.
We will continue this section by describing different hadronization-$p_\perp$ models, which give slightly different predictions on $Q_\perp$ and the other observables discussed in section~\ref{sec:obs}. In section~\ref{sec:results} we examine how well these differences can be distinguished in a two-jet event sample.

\subsection{J{\small ETSET} default}\label{sec:default}
In the \textsc{Jetset} default version of the string
fragmentation model the \pair q breakups of the string are
uncorrelated both in transverse momentum and in flavour. As the quark
and the antiquark in a pair have balancing $k_\perp$ the transverse momentum of one hadron is fully compensated by its two neighbours in rank. This implies that $p_\perp$ is conserved very locally in rapidity.

From Eq~(\ref{e:tunnel})  we expect that the width of the $k_\perp$-distribution should be
given by $\sqrt{\kappa/\pi} \approx 250$ MeV, but in applications it is
treated as a tunable parameter, correlated to the cut-off for the parton cascade. Fits to data give a value around 360 to 400
MeV. 
Eq~(\ref{e:tunnel}) also introduces an overall suppression of heavy quarks. This implies that the rate of c is completely negligible (the c/u ratio is $\sim 10^{-11}$). 
To get the rate of s-quarks we would have to know its
effective mass. The observed s/u ratio 
around 1/3 corresponds to the very reasonable value 250 MeV.

Studies of photons at LEP1 indicate a larger tail for high-$p_\perp$
$\pi^0$s~\cite{pi0pt}. The origin of this tail is not clear, but in \textsc{Jetset} default a
small component with higher $p_\perp$ is included ($k_\perp$ is multiplied
by a factor of 2 for 1\% of the vertices).

\subsection{Partial $p_\perp$ Compensation}\label{sec:jim}

The production rate for $\eta$ and $\eta^\prime$ cannot be reproduced if
neighbouring string breakups are uncorrelated in flavour. Also
studies of two-particle correlations observed in hadron--hadron collisions are not
well described if the ${\mathbf k}_\perp$ in different breakups are totally
uncorrelated~\cite{K2}. As discussed above, this would imply that the $p_\perp$ of a
hadron is fully compensated by its two neighbours in rank. In ref~\cite{jim} a
model is presented in which there are correlations in flavour and $k_\perp$
with some finite correlation length. The model implies a suppression of  $\eta$ and $\eta^\prime$ production, and also that only a fraction $\gamma<1$ of the $p_\perp$ of a hadron is compensated by the neighbouring hadrons in rank. Thus $p_\perp$ is compensated within a finite correlation length, but this range is larger than in \textsc{Jetset} default.

In~\cite{jim} it is shown that this $p_\perp$ compensation assumption can be embedded in an iterative fragmentation scheme. The inclusive $p_\perp$ distribution is then a Gaussian,
\eqbe \mrm dP\propto \exp(-\frac{\alpha}{2\gamma}p_\perp^2)\mrm d^2{\mathbf p}_\perp, \label{e:jimpt} \eqen
where $\alpha$ is a tunable parameter. Furthermore, the sum in ${\mathbf p}_\perp$ for the first $i$ particles in rank, ${\mathbf Q}_{\perp i}$, saturates for large $i$ towards the distribution
\eqbe \mrm dP\propto \exp(-\alpha Q_{\perp}^2)\mrm d^2{\mathbf Q}_{\perp}. \label{e:jimQt} \eqen
We note that for fixed $\left<p_\perp^2\right>$ for the hadrons, the collective variable $\left<Q_{\perp i}^2\right>$ is smaller for larger $\gamma$. This is just the previously stated behaviour for the closely related observable  $\left<Q_{\perp}^2(y)\right>$, which is small if $p_\perp$ is conserved locally in rapidity.

The \textsc{Jetset} default algorithm is retrieved by setting $\gamma=1$ and $\alpha=1/\sigma^2$, where $\sigma$ is the width in Eq~(\ref{e:tunnel}). 
In~\cite{jim}, $\gamma$ is instead assumed to depend on the hadron mass $M$, and is parametrized as 
\eqbe \gamma=\frac1{1+M_0/M}, \label{e:M0def} \eqen
where $M_0$ is set to $0.9$ GeV.

\subsection{Uncorrelated $p_\perp$}
To test whether the data really show a local $p_\perp$ conservation, we also want to study a situation where the hadronic transverse momenta are uncorrelated apart from overall $p_\perp$ conservation. This can be achieved by pushing $M_0$ to infinity in Eq~(\ref{e:M0def}), which corresponds to $\gamma$ being small, but keeping $\alpha/\gamma$ finite. The MC from~\cite{jim} is not designed to cope with very large $M_0$-values. When hadrons are peeled off from both ends of the string usually a large $p_\perp$ is built up for a remaining string piece in the centre. Such events have to be discarded, keeping only those events where by chance this string piece has limited $p_\perp$. In this paper we study only central particles in the accepted events, and thus for our purposes we obtain a fair estimate of the result if in the MC we peel off hadrons from one end only, pushing the junction to the other and outside the region of study.

\subsection{The UCLA model}\label{sec:ucla}

A partial $p_\perp$ compensation is also assumed in the UCLA model~\cite{ucla}. In this
model it is assumed that not only the longitudinal momentum distribution
but also the flavour composition and the distribution in transverse
momentum are determined by the symmetric fragmentation function in Eq~(\ref{e:fz}). Thus in this model the quark degrees of freedom do not appear
explicitly, but only the momenta and flavour of the hadrons. A direct
application of Eq~(\ref{e:fz})  will however give a too wide $p_\perp$-distribution.
In the UCLA approach this problem is solved by assuming a similar kind
of finite $p_\perp$ correlations as in the model in section~\ref{sec:jim}, with $\gamma=1/2$. Thus as far as transverse momentum goes we expect rather similar results for the UCLA model and the model in~\cite{jim}, and for this reason we content ourselves to study the latter in our investigatons.

\subsection{Helix String}\label{sec:helix}
At the end of a perturbative parton cascade the virtualities become
relatively small and the running coupling therefore relatively large.
This implies that interference and coherence effects are expected to be
important. In ref~\cite{screw} arguments are presented for correlations between
rapidity and azimuthal angle for the emitted gluons, which correspond to
a helix-like field at the end of the cascade. This field
configuration has also some similarities with a string coupled to a
gauge field or a massive relativistic string~\cite{stringandfield}.

A modification to the Lund fragmentation scheme reflecting the helix-like field is presented in~\cite{screw}. The rapidity distance $\Delta y$ between two string breakups is correlated with the azimuthal angle $\Delta\phi$ between the transverse direction of the two produced \pair q pairs. The pitch of the helix is given by an unknown parameter $\Delta y/\Delta\phi=\tau$, allowing for some Gaussian fluctuations around this value. This implies that the observable $S(\omega)$ defined in section~\ref{sec:obsscrew} gets a peak at $\omega\approx 1/ \tau$. Expected values of $ \tau$ are around 0.3 to 0.5. The string energy per unit rapidity available to generate $p_\perp$ is given by a parameter $m$, which implies that a typical magnitude of hadron $p_\perp$ is given by $m \tau$.

\subsection{Time Dependent $p_\perp$} \label{sec:timept}
In order to investigate the sensitivity of the observables we will also study a model with a correlation between transverse momentum and multiplicity which might not be totally inconceivable. The results in Eq~(\ref{e:tunnel}) with a constant width corresponds to a situation with a static flux tube or stringlike colour field. The estimated width of this flux tube is of the order of 1 fm. One can imagine that at very early times the flux is more concentrated. This might correspond to a larger effective string tension, and also from the uncertainty relation one might then imagine that larger transverse momenta are produced in early breakups. If the transverse dimension of the flux tube equals the proper time $\tau$, then we could expect fluctuations in $k_\perp$ of the order $2/\tau$. Expressed in the variable $\Gamma=\kappa^2\tau^2$ this corresponds to a width $\sigma^2\sim 4\kappa^2/\Gamma$. This becomes very large when $\Gamma\ra 0$, and from a string picture we should expect a saturation. To produce large $k_\perp$ sufficient energy must be stored in the string, which therefore must have a minimum length. As discussed in~\cite{tunnelhyperbola} we expect a suppression when $k_\perp>\kappa\tau$, and thus a saturation when $2/\tau\sim \kappa\tau$ or $\Gamma\sim 2\kappa$. These arguments would give the following width for the $k_\perp$-generation
\eqbe \sigma^2=\left\{\begin{array}{ll} \frac{\kappa}{\pi}+\frac{4\kappa^2}{\Gamma}, & \Gamma<2\kappa \\ \sigma\srm{max}^2=\frac{\kappa}{\pi}+2\kappa, & \Gamma>2\kappa \end{array} \right. \eqen
Actually this expression does reproduce the inclusive $p_\perp$ spectrum with hardly any tuning. In our calculations we allowed for an overall normalization factor on $\sigma$,
\eqbe \sigma\ra\chi\sigma. \eqen
For the standard perturbative cutoff we get $\chi=1.08$.

A special consequence of the time dependent $k_\perp$ width is a strong correlation between transverse momentum and multiplicity. An early breakup splits the original system in two pieces with relatively small invariant masses. If these masses are called $M_1$ and $M_2$ we have approximately 
\eqbe M_1^2M_2^2\approx\Gamma s, \eqen
where $\sqrt s$ is the total mass of the system. It is also easy to show that the ensuing depletion in multiplicity appears in a rapidity  of size $\Delta y\sim\ln(\left<\Gamma\right>/\Gamma)$ around the breakup, where $\left<\Gamma\right>$ denotes the average value of $\Gamma$.

\section{Results}\label{sec:results}
\subsection{Hadron $p_\perp$ spectra}
\begin{table}[tb]
  \begin{center} 
    \begin{tabular}{|l||l|l|l||l|l|l||l|l||}
\hline
model & \multicolumn{6}{l||}{parameters}  & $\left<N\right>$ & $\left<p_\perp\right>$\\
& $k_{\perp\mrm{cut}}$ & $a$ & $b$ & \multicolumn{3}{l||}{others} & & \\
& (GeV) & &  (GeV$^{-2}$) & \multicolumn{3}{l||}{} &  & \\
\hline
Default & & & & $\sigma_{k_\perp}$(GeV) & & & &\\
\cline{5-5}
 & $0.6$ & $0.23$ & $0.34$ & $0.405$ &  & & $34.9$  & $0.60$ \\
 & $1.5$ & $0.5$ & $0.4$ & $0.42$ &  &  &  $35.1$ & $0.59$\\
\hline
Partial $p_\perp$ & & & & $\frac{1}{\sqrt{\alpha}}$(GeV) & $M_0$(GeV) & & &\\
\cline{5-6}
Compens.\ & $0.6$ & $0.23$ & $0.34$ & $0.67$ & $0.9$ &  & $34.9$ &  $0.60$\\
\hline
Uncorr.\ $p_\perp$ & $0.6$ & $0.23$ & $0.34$ & $1.52$ & $10$ &  & $34.8$ &  $0.59$\\
\hline
Helix & & & & $ \tau$ & $\sigma_\tau$ & $m$(GeV) & &\\
\cline{5-7}
 & $0.6$ & $0.23$ & $0.37$  & $0.3$  & $0.2$ &  $1.1$ & $35.0$ &  $0.59$\\
 & $0.6$ & $0.23$ & $0.39$  & $0.5$ & $0.3$ & $0.8$  & $34.8$ & $0.60$ \\
 & $0.6$ & $0.23$ & $0.4$  & $0.7$& $0.35$ & $0.68$ & $34.6$ &  $0.60$\\
\hline
Time & & & & $\chi$ & & & &\\
\cline{5-5}
Dep. $p_\perp$ & $0.6$ & $0.23$ & $0.34$  & $1.08$ & &   & $34.7$ & $0.60$ \\
 & $1.5$ & $0.5$ & $0.42$  &  $1.18$ & &  & $34.7$ &  $0.60$\\
\hline
    \end{tabular}
  \caption{\em The tuned parameter values of the considered models. The tunes are performed w.r.t.\ average multiplicity $\left<N\right>$, with $\pi^0$ stable, and average $\left<p_\perp\right>$ within a rapidity interval $|y|<y\srm{cut}=3$. $\left<p_\perp\right>$ is measured w.r.t.\ thrust axis and defined as $\frac 1{n\srm{ev}\left<N(y\srm{cut})\right>}\sum\srm{ev}\sum_{|y|<y\srm{cut}}\left|p_\perp\right|$. The top line of the models is the \textsc{Ariadne}$+$\textsc{Jetset} default tune which has not been changed. The other models are tuned to give similar results. The models and their parameters are discussed in section~\protect\ref{sec:models}.}
  \label{tab:tunes}
  \end{center}
\end{table}
We have used average $p_\perp$ and average multiplicity to impose constraints on the free parameters of the considered models. For simplicity, we have left the \textsc{Ariadne}+\textsc{Jetset} default algorithm unchanged and tuned the other models to reproduce its values. Differences are then searched for in other observables, which will be discussed below.

All models depend on the $k_{\perp\mrm{cut}}$ of the cascade and the $a$ and $b$ parameters of the symmetric fragmentation function Eq~(\ref{e:fz}). To compare different hadronization models, we are interested in using the same description of the perturbative phase for all models, and we have therefore tuned all models  keeping the cascade cut-off $k_{\perp\mrm{cut}}$ constant to the \textsc{Ariadne} default value. As $k_{\perp\mrm{cut}}$ is also  an hadronization parameter, we will in some examples compare tunes of the same model with different values of $k_{\perp\mrm{cut}}$.

The $a$ and $b$ parameters influence the multiplicity distribution. For each setup of other parameters, there is in general a narrow but long band of values in the $a$ and $b$ parameter space which reproduces the average multiplicity. The specific $a$ and $b$ values in the tunes are chosen to give a fair agreement also on multiplicity dispersion. The result is presented in Table~\ref{tab:tunes}.

\subsubsection*{$p_\perp$ in low multiplicity events}
The different models have been tuned to give similar $p_\perp$ distributions in the full event sample. After a two-jet cut, the $p_\perp$ distribution may differ, as the importance of the perturbative phase is then reduced. We will here investigate that possibility. Among the two-jet cuts discussed in section~\ref{sec:cuts}, the rather blunt cut in multiplicity is then of interest, as it introduces little bias to the $p_\perp$ spectrum.

In section~\ref{sec:timept}, we discussed a model where the proper time of a \pair q vertex in the string influences the width of the $k_\perp$ distribution for the \pair q pair. This model introduces correlations between average $p_\perp$ and multiplicity.

Gluon emissions in general contribute  both to $p_\perp$ and multiplicity. For soft gluons the effects are smaller, but if their contribution to $p_\perp$ dominates, it affects the average $p_\perp$ in low-multipicity events. Thus it is of interest to compare correlations in $\left<p_\perp\right>$ and multiplicity for the same fragmentation model tuned with different cascade cutoffs.

\begin{figure}[tb]
  \begin{center}  \hbox{ \vbox{
	\mbox{\psfig{figure=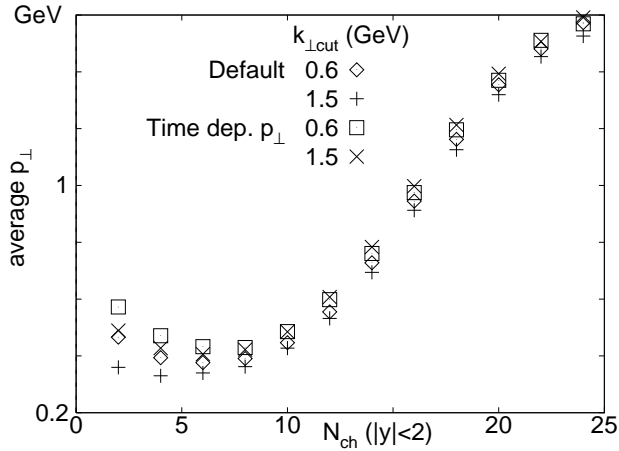,width=\figwidth}}
   }  }
  \end{center}  \caption{\em Average $p_\perp$ w.r.t.\ thrust axis for different charged multiplicities. Statistical errors are well within symbol sizes. The correlation between $p_\perp$ and multiplicity in the time dependent $p_\perp$ model is seen as a higher prediction for average $p_\perp$ in low multiplicity events. For both models $\left<p_\perp\right>$ is higher in low multiplicity events when the cascade cutoff $k_{\perp\mrm{cut}}$ is lowered. This indicates that soft gluons in these models give a larger contribution to hadron $p_\perp$ than to hadron multiplicities.}
  \label{f:ptnch}
\end{figure} 
In Fig~\ref{f:ptnch} we compare the time dependent $p_\perp$ model discussed in section~\ref{sec:timept} and  \textsc{Jetset} default fragmentation by studying average $p_\perp$ of particles in the central rapidity region $|y|<2$, as a function of the charged multiplicity $N\srm{ch}$ in that region. As expected, average $p_\perp$ at low multiplicities is larger in the  time dependent $p_\perp$ model. $p_\perp$ and $y$ are defined w.r.t.\ the thrust axis, and no two-jet cut in a $p_\perp$ measure is imposed on the event sample.

We note that  $\left<p_\perp\right>$ at low multiplicities is higher for both models if the cascade cutoff is reduced from 1.5 GeV to 0.6 GeV. In the dipole cascade combined with string fragmenation, the soft gluons in this range of $k_\perp$ thus contributes to hadron $p_\perp$ more than to hadron multiplicities.

We have also checked the prediction for the time dependent $p_\perp$ model if the expected saturation of the \pair q $k_\perp$ width at very early breakups is neglected. We then get a significantly higher prediction for $\left<p_\perp\right>$. When  $N\srm{ch}(|y|<2)=2$, it lies  between 0.9 and 1.2 GeV, depending on the $k_{\perp\mrm{cut}}$ for the cascade. The correlation in $p_\perp$ and multiplicity is thus and observable which can be used to confront this kind of unexpected $p_\perp$ generation with experimental data.

\subsection{$p_\perp$ transfer}
The observable $Q_\perp(y)$ discussed in section~\ref{sec:obsQt} measures the $p_\perp$ transfer over the rapidity $y$ and is sensitive to correlations in hadron $p_\perp$. We have investigated $Q_\perp$ for different assumptions about hadron $p_\perp$ conservation. In the default algorithm, $p_\perp$ is compensated by the neighbours in rank, which implies a compensation local in rapidity. In the partial $p_\perp$ compensation model, a fraction $\gamma<1$ of $p_\perp$ is compensated by closest neighbours. This implies a correlation length for $p_\perp$ of order $1/\gamma$ which is finite, but larger than for \textsc{Jetset} default. We also examine a model where the hadron $p_\perp$ are uncorrelated, apart from global momentum conservation. 

In two-jet events, the hadrons can also aquire $p_\perp$ from soft gluons. The compensation of this $p_\perp$ is in general not identical to the one assumed in the fragmentation, and depends on the recoil treatment of the cascade formalism. Thus the assumed $p_\perp$ conservation length may depend on the treatment of recoils in the cascade, and on the cutoff scale, which determines to what extent hadron $p_\perp$ originates from the cascade or the hadronization.

\begin{table}[tb]
  \begin{center} 
    \begin{tabular}{|ll||l|l|l|l||}
\hline
\multicolumn{2}{|c||}{$\left<Q_\perp^2\right>$(GeV$^2$)} & \multicolumn{2}{c|}{default} & partial $p_\perp$ & uncorr.\ $p_\perp$ \\
& & \multicolumn{2}{c|}{ } & compens. &\\
\hline
& $k_{\perp\mrm{cut}}$ (GeV)  & 1.5 & 0.6 & 0.6 & 0.6\\
\hline
No cascade & & - & 0.60 & 0.71 & 0.74 \\
With cascade, & all events & 12.2 & 12.3 & 12.4 & 12.6 \\
 & two-jet events & 0.86 & 0.94 & 1.02 & 1.09\\
\hline
    \end{tabular}
  \caption{\em Average $Q_\perp^2$, measured w.r.t.\ the sphericity axis in a central rapidity range $|y|<2$. Results for different assumptions about hadron $p_\perp$ correlation lengts. The differences in the predictions are shadowed by the perturbative cascade, but can be restored by a two-jet cut. Here a cut in sphericity, $\ln(S)<-4.5$, has been used.}
  \label{tab:QtQt}
  \end{center}
\end{table}
The top row in Table~\ref{tab:QtQt} shows results without cascade for the three different $p_\perp$ correlation assumptions. $Q_\perp$ clearly grows for less local $p_\perp$ compensation, with a relative difference of 25\% between complete $p_\perp$ compensation by neighbours (\textsc{Jetset} default) and uncorrelated $p_\perp$. Adding a cascade allmost wipes out the difference, but after a two-jet cut it is restored to a satisfactory degree. The relative difference between the two extremes with a common cutoff $k_{\perp\mrm{cut}}$$=0.6$ GeV (second and fourth column) is then about 16\%. Also partial $p_\perp$ correlation as assumed in~\cite{jim}, similar to the UCLA $p_\perp$ generation model, gives results after a  proper two-jet cut which differ significantly from the results of the \textsc{Jetset} default assumption.

A comparison of the first and second column illustrates the difference in $p_\perp$ compensation in the cascade and in the fragmentation. The same $p_\perp$ model, tuned with different cascade cut-offs, give significantly different results on $Q_\perp$. Soft gluon emission in the dipole cascade introduces longer $p_\perp$ correlation lengths than given by default hadronization $p_\perp$, which implies that $Q_\perp$ increases when adding soft gluons in the range 1.5 to 0.6 GeV to the cascade. 

\subsection{Screwiness}
In~\cite{screw}, MC simulations of \pair q string fragmentation without cascade show a clear signal in $S(\omega)$ in the helix-like fragmentation model for $ \tau\sim 0.5$ or larger. To check the influence of relatively soft gluons on the screwiness measure, we have modified \textsc{Jetset} to generate $p_\perp$ according to the helix model. The $p_\perp$ generation is then  combined with the \textsc{Jetset} default recipe for passing gluons~\cite{gluonkinks}. 
This implies that the azimuthal directions on different sides of a gluon kink is not completely uncorrelated, though measured w.r.t.\ two different directions. The $p_\perp$ component being transverse to both considered directions is preserved when passing a gluon kink.

One could also consider more direct effects, assuming  the azimuthal angle of vertices on different string segments to be explicitly uncorrelated, and also allowing the helix to have different orientation on different string segments. The investigation here however focus on the minimal effects of gluons bending the string. 

\begin{figure}[tb]
  \begin{center}  \hbox{ \vbox{
	\mbox{\psfig{figure=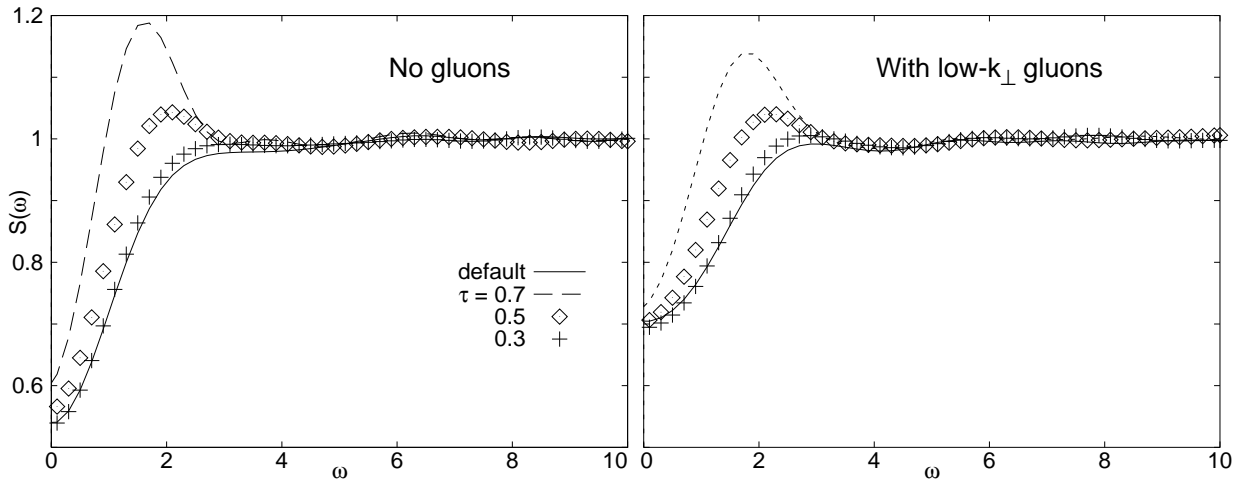,width=\textwidth}}
   }  }
  \end{center}  \caption{\em Screwiness without gluons, and with low-$k_\perp$ gluons ($k_\perp$ in the emissions is restricted between 1 and 2 GeV). The central rapidity region $|y|<2$ is considered. With the assumptions made here about effects from perturbative gluons, the signal survives.}
  \label{f:gscrew}
\end{figure} 
Before looking at screwiness after a full cascade and a two-jet cut, we examine the effects of soft gluons by running a cascade with $k_\perp$ for the emissions constrained between 1 and 2 GeV. As seen in Fig~\ref{f:gscrew}, the observable $S(\omega)$ is fairly unaffected by these rather soft gluons. 

\begin{figure}[tb]
  \begin{center}  \hbox{ \vbox{
	\mbox{\psfig{figure=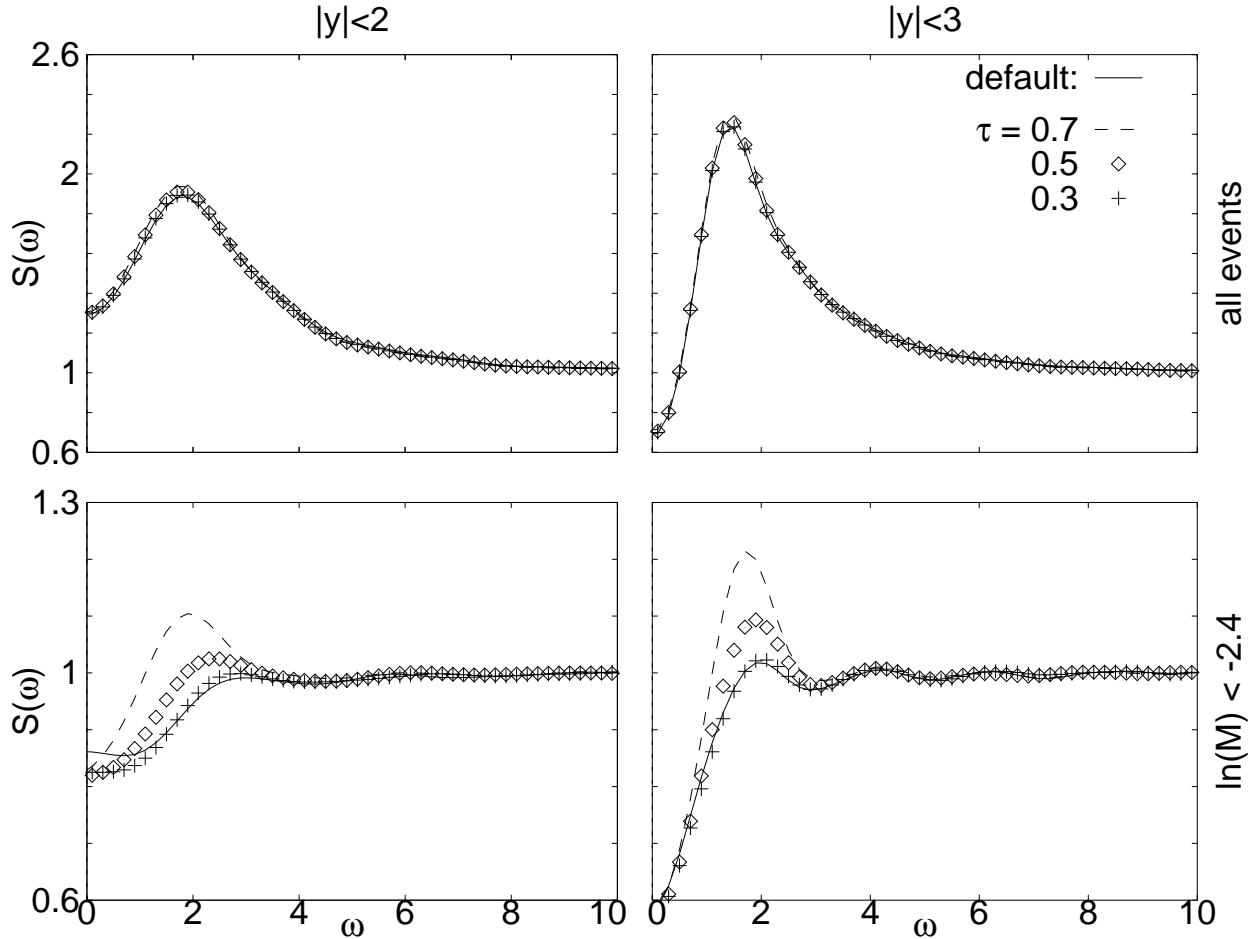,width=\textwidth}}
   }  }
  \end{center}  \caption{\em Screwiness after gluon cascade, with and without a two-jet cut. Without any two-jet cuts (upper row), a signal reflecting long-range azimuthal correlations in three-jet structures is seen, independent of fragmentation mechanism. After a two-jet event cut, this signal is removed, and the $S(\omega)$ measure is sensitive to helix-like fragmentation. The results are similar to the ones with only low-$k_\perp$ gluons, and only moderately reduced.}
  \label{f:cutscrew}
\end{figure} 
A cascade introduces long-range azimuthal correlations which have large effects on $S(\omega)$. In a three-jet event the thrust direction tends to be aligned with the hardest jet, which implies that the particles of this jet are rather randomly distributed in azimuthal angle $\phi$. The softest jet is seen as a set of particles well collimated both in $\phi$ and $y$. The particles of the remaining jet are more spread out towards large rapidities and have and azimuthal direction opposite to the softest jet. In this typical three-jet configuration, particles of the hardest jet merely introduces noise to $S(\omega)$, while the two softer jets tend to enhance $S(\omega)$ when $\omega\sim \pi/y\srm{cut}$, where $y\srm{cut}$ is the largest considered rapidity in the analysis. Thus we get a signal which is independent of fragmentation mechanism, but is sensitive to the magnitude of the considered central rapidity range. This behaviour is seen in the upper plots of Fig~\ref{f:cutscrew}.

After event cuts (lower plots in Fig~\ref{f:cutscrew}), the result is very similar to those with low-$k_\perp$ gluons in Fig~\ref{f:gscrew} and give a clear signal for helix-like string fragmentation if $ \tau\sim 0.5$ or larger. In~\cite{screw}, different ways to enhance the signal also for lower values of $ \tau$ were investigated. The analysis was performed on \pair q strings without cascade, and it was shown that a lower limit on average $p_\perp$ could be used to distinguish default string fragmentation and helix fragmentation with $ \tau$ down to 0.3. A lower cut in $p_\perp$ is slightly in conflict with a two-jet cut, but it is possible that a similar approach can be used to enhance the signal in two-jet events obtained from real data.

\section{Conclusions}~\label{sec:concl}
High energy reactions like $e^+e^-$$\ra$ hadrons are usually described in terms of two phases, an initial perturbative phase, formulated as a parton cascade, followed by a soft hadronization phase described by a phenomenological model. It is however not only the description of the soft phase which depends on model assumptions. This is also the case for the perturbative phase towards the end of the cascade, where $\alpha_s$ is large and some kind of cut-off is necessary. The transition region may also exhibit interesting interference or coherence effects, one example being the helix-like field configuration proposed in~\cite{screw}. 

In many cases the effects of gluon emission overshadow the features of the hadronization phase. To investigate the hadronization mechanism it is therefore interesting to study whether suitable events cuts can help separating the two phases and give observables sensitive mainly to one of them.

We have examined different strong two-jet cuts to suppress the perturbative activity. It is possible to define efficiency and purity measures for the different cuts. MC simulations show, however, that no cut can be called optimal. Instead the preferred cut depends on the observable under investigation.

As the flavour composition of the final state in most models is less sensitive to the parton cascade, we concentrate here on observables related to transverse momentum. Different assumptions can be made concerning recoil effects and the locality of $p_\perp$ conservation. In most models it is possible to tune the inclusive $p_\perp$ spectrum, and to distinguish the models it is then necessary to study different forms of correlations. The vectorial sum of ${\mathbf p}_\perp$ for all particles with rapidities smaller than some value $y$, here called ${\mathbf Q}_\perp$, is an observable which is sensitive to the $p_\perp$ correlation length. If $p_\perp$ is locally conserved, average $Q_\perp$ is similar to average $p_\perp$. On the other hand, if only global constraints restrict $p_\perp$, the collective measure $Q_\perp$ gets significantly larger than $p_\perp$.

After a two-jet cut, we find the $Q_\perp$ measure to be sensitive to assumptions about $p_\perp$ compensation in the hadronization phase. It is also sensitive to the value of the cascade cut-off, since the recoil treatment in the cascade in general give longer $p_\perp$ correlation lengths than assumed in the fragmentation.

Another observable which is shown to be sensitive to the cascade cut-off is average $p_\perp$ in low multiplicity events. Though multiplicity in general is not considered a two-jet cut, it suppresses high-$k_\perp$ perturbative emissions, which typically give rise to large multiplicities. Gluon emissions with moderate $k_\perp$ however contribute more to hadronic $p_\perp$ than to hadronic multiplicity, which implies that the correlation between $p_\perp$ and multiplicity is sensitive to the cascade cut-off.

In~\cite{screw} it is argued that coherence effects in the transition region may give rise to a helix-like colour field, with a correlation beteen rapidity and azimuthal angle. Perturbative gluon emissions mask this correlation to a large extent, but we have found that after a strong two-jet cut, a signal in the proposed observable ``screwiness'' is present for some possible helix configurations.

The very high statistics now available from experiments at the ${\mrm Z}^0$ resonance implies that strong two-jet cuts are possible on the data without loosing the statistical significance. Our results show that such cuts can indeed in many cases discriminate between different hadronization models and thus be a tool for a more detailed study of the hadronization mechanism.

\end{document}